\documentclass[aps,prl,preprintnumbers,a4,showpacs,twocolumn,nofootinbib]{revtex4}

\newcommand{\be}{\begin{eqnarray}}
\newcommand{\ee}{\end{eqnarray}}

\def\hbar#1{\slash\hspace{-2.5mm}#1}
\usepackage{graphicx,graphics}% Include figure files
\usepackage{mathrsfs}%for slanted letters
\usepackage{amssymb,amsmath}
\usepackage{dcolumn}% Align table columns on decimal point
\usepackage{bm}% bold math
\usepackage{slashed}%allow feynman slash notation with \slased{} command;
\usepackage{color}

\newcommand{\beq}[1] {\begin{equation}\label{#1} }
\newcommand{\eeq} {\end{equation} }
\newcommand{\bea}[1]{\begin{eqnarray}\label{#1} }
\newcommand{\eea}{\end{eqnarray}}

\begin{document}

\preprint{OSU-HEP-10-07}
\title{Non-renormalizable Yukawa Interactions and Higgs Physics}
\author{Z.~Murdock}
%\email{zeke.murdock@okstate.edu}
\author{S.~Nandi}%\email{s.nandi@okstate.edu}
\author{Santosh Kumar Rai}%\email{santosh.rai@okstate.edu}
\affiliation{Department of Physics and Oklahoma Center for High Energy Physics,
Oklahoma State University, Stillwater, Oklahoma, 74078.}%

\begin{abstract}
We explore a scenario in the Standard Model in which dimension four
Yukawa couplings are either forbidden by a symmetry, or happen to be
very tiny, and the Yukawa interactions are dominated by effective
dimension six interactions. In this case, the Higgs interactions to
the fermions are enhanced in a large way, whereas its interaction
with the gauge bosons remains the same as in the Standard Model. In
hadron colliders, Higgs boson production via gluon gluon fusion
increases by a factor of nine. Higgs decay widths to fermion
anti-fermion pairs also increase by the same factor, whereas the
decay widths to photon photon and $\gamma Z$ are reduced.
Current Tevatron exclusion range for the Higgs mass increases to
$\sim 142- 200$ GeV in our scenario, and new physics must appear at
a scale below a TeV.
\end{abstract}
%%%%%%%%%%%%%%%%%%%%%%%%%%%%%%%
\maketitle
%%%%%%%%%%%%%%%%%%%%%%%%%%%%%%%
%\section{Introduction}

The Standard Model (SM) based on the gauge symmetry $SU(3)_C \times
SU(2)_L \times U(1)_Y$ is in excellent agreement with all the
current experimental results. However, there are sectors of the SM
which are still untested, such as the Higgs sector and the Yukawa
sector. In the SM, we have only one Higgs doublet, and we allow the
Higgs self interactions up to dimension four to maintain the
renormalizability of the theory. In this case, the cubic ($h^3$) and
the quartic ($h^4$) interactions of the remaining neutral scalar
Higgs field, $h$ is determined in terms of the Higgs mass, $M_h$ and
the known vacuum expectation value (VEV), $v$. Although we know $v$
experimentally to a very good accuracy, the Higgs mass is still
unknown. Hence its presence, as well as the magnitude of its cubic
and quartic self interactions are completely untested. The other
untested sector of the SM is the Yukawa sector. In the SM, we
introduce dimension four Yukawa interactions which give masses to
the fermions, and also generate the Yukawa interactions between the
Higgs field $h$ and the fermions. The strength of these Yukawa
interactions are completely determined in terms of the fermion
masses and $v$. However, we do not have any experimental evidence
for these interactions being the source of the fermion masses, and
the presence of these dimension four Yukawa interactions. Another
point to emphasize is that we do not know whether the Higgs boson is
elementary or composite. Theories have been formulated in which the
Higgs boson is a fermion anti-fermion composite; or more specifically
a condensate of the third family quark and anti-quark \cite{Miransky:1989ds}.
Other possibilities for composite Higgs have also been
advocated \cite{Hill:1990ge,Hill:2002ap}.
Whether the Higgs boson is an elementary particle or composite, the
operators of dimension higher than four suppressed by some scale,
$M$ are expected. It has also been pointed out that the presence of
dimension six operator in the Higgs potential allows us to have
baryogenesis via sphaleron \cite{Grojean:2004xa}, still satisfying the
current LEP limit on the Higgs mass.

In this letter, we propose an alternate scenario for the Yukawa
sector, and explore how to test our predictions experimentally at
the Tevatron and LHC. The effects of general dimension six operators
in the Higgs sector have been considered and studied before
\cite{Barger:2003rs}. Also other dimension six operators may appear
in SM and a complete list of such operators is collected in Ref.
\cite{Buchmuller:1985jz}. We consider the case in which the usual
dimension four Yukawa interactions are either forbidden by a
symmetry, or the corresponding coupling happens to be too tiny to
generate the observed values of the fermion masses. In this case,
the dominant contribution to the fermion masses, as well as the
interactions between the fermions and the Higgs boson will arise
from the dimension six effective Yukawa interactions of the form
$(f/M^2)\bar{\psi_L} \psi_R H (H^{\dag} H)$, where $M$ is the mass
scale for the new physics through which such effective interactions
are generated. As in the SM, fermion masses are still parameters in
the theory, but the Yukawa couplings of the fermions to the Higgs
boson are a factor of three larger than the SM. This enhances the
production of the Higgs boson, as well as affect its decay
branching ratios to various final states. This will have interesting
consequences for Higgs signals at the Tevatron and LHC, as well as
in the possible future lepton collider.
%%%
%\section{Formalism}
%%%%

Our model is based on the SM gauge symmetry, $SU(3)_C \times SU(2)_L
\times U(1)_Y$. We denote the left handed electroweak (EW) quark
doublets by $q_{Li} \equiv (u,d)^T_{Li}$, and the right handed EW
quark singlets by $u_{Ri}$ and $d_{Ri}$, where the index $i$ ($i=
1,2,3$) represent three fermion families. Then the Yukawa
interactions of the fermions with the Higgs boson up to dimension
six are given by
%%%
\begin{align}
& \mathcal{L}_{\text{Yukawa}} = \bar{q}_L f_u u_R \widetilde{H} +
\bar{q}_L f_d d_R H + \bar{l}_L f_L e_R H \nonumber \\
%\mathcal{L}_{\text{Yukawa}}
+ & \frac{1}{M^2}( \bar{q}_L y_u u_R \widetilde{H} + \bar{q}_L y_d
d_R H + \bar{l}_L y_L e_R H)(H^{\dag} H) + h.c.,
\label{yuk}
\end{align}
where the fermion fields represent three families, and  $f_d, f_u$
and $f_l$ represent three corresponding Yukawa coupling matrices for
the dimension four Yukawa interaction while $y_d, y_u$ and $y_l$
represent three corresponding Yukawa coupling matrices for the
dimension six Yukawa interactions. $M$ is the mass scale for a new
physics which generates these dimension six interactions.

Our proposed scenario is the case in which the dimension four Yukawa
couplings, $f_d, f_u$ and $f_l$ are either forbidden by a symmetry,
or happen to be very tiny to generate the observed fermion masses,
and this sector is dominated by dimension six interactions given
above. Thus, choosing the couplings $f$ to be zero, for the
fermion mass and the Yukawa coupling matrices, we obtain
%%%%
\begin{align}\label{new}
\mathcal{M}_{\text{New}} &= \frac{1}{2 \sqrt{2}M^2} y_d (v^3),
\nonumber \\
\mathcal{Y}_{\text{New}} &= \frac{1}{2 \sqrt{2}M^2} y_d (3 v^2),
\end{align}
%%%
and similar expressions for the up quark and lepton sector. In
contrast, in the usual SM, where we do not include the effective
dimension six interactions, we have
\begin{align}\label{SM}
\mathcal{M}_{\text{SM}} = \frac{1}{\sqrt{2}} f_d (v), &&
%\nonumber \\
\mathcal{Y}_{\text{SM}} = \frac{1}{ \sqrt{2}} f_d.
\end{align}
%%%%
In our scenario, one can see from Eq.~\ref{new} that the mass
matrices and the corresponding Yukawa coupling matrices are
proportional. Hence as in the usual SM, we do not have any Higgs
mediated flavor changing neutral current interactions. The important
point to note is that in our scenario (for simplicity, we call it the
new model), the Yukawa couplings
of the Higgs boson to the fermions are three times larger than those
in the SM, whereas the gauge interaction of the Higgs boson remains
the same. This will make important differences for Higgs production,
and its decay branching ratios as we discuss below.
%%%%
\begin{widetext}
\begin{center}
\begin{figure}[!ht]
\includegraphics[width=7.0in,height=2.5in]{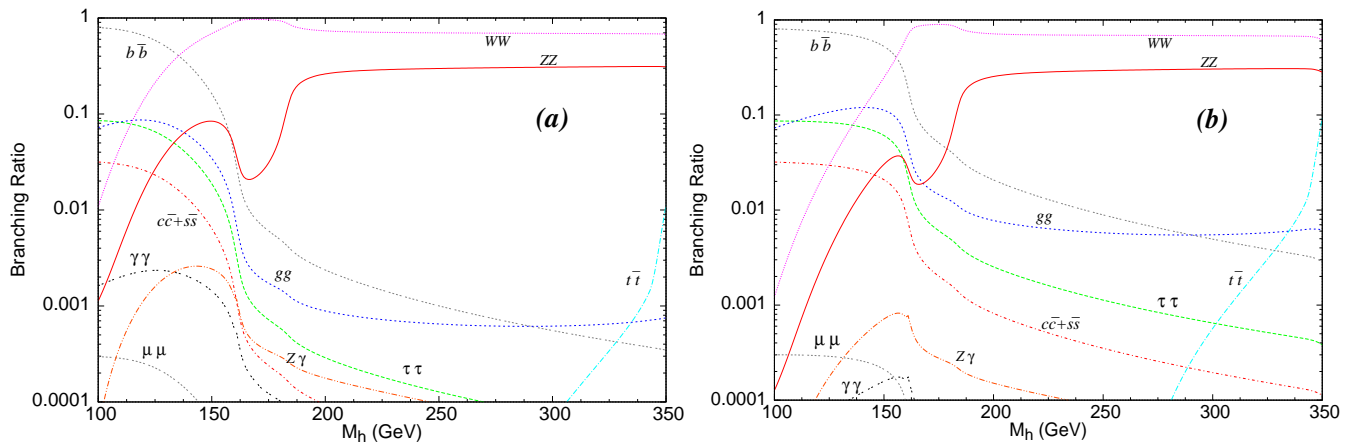}
\caption{Illustrating the branching ratios for Higgs decays in (a) SM and (b) new model
as a function of its mass. We have used the package {\sc Hdecay} \cite{hdecay} to calculate
the Higgs decay modes.}
\label{h-2x}
\end{figure}
\end{center}
\end{widetext}
%%%%%
%\section{Phenomenological implications}
%%%
%\subsection{Higgs decays}
%%%%
In the low Higgs mass range ($M_h \leq 125$ GeV), the Higgs boson
dominantly decays to $b \bar{b}$ in the SM. This mode is even more
dominant in the new model, since the $h b \bar{b}$ coupling is
enhanced by a factor of three compared to the SM. In the SM, the $b
\bar{b}$ to $WW$ crossover takes place at $M_h \sim 135$ GeV (see
fig. \ref{h-2x}a), while in our model, this crossover happens at
$M_h \sim 155$ GeV, (see fig. \ref{h-2x}b). Also, as can be seen from
these figures, the $\gamma \gamma$ branching fraction in our model
is suppressed by about a factor of ten compared to the SM.
The reason is that in the $h \rightarrow \gamma \gamma$ decay, the
contribution comes from the $W$ loop and the top quark loop, and the
two contributions are of opposite sign. In our model, because the $h
t \bar{t}$ coupling is enhanced by a factor of three, there is a
strong cancelation between the top loop and the $W$ loop
contributions, resulting in the large suppression in the $\gamma
\gamma$ mode. Note that in our model, Higgs couplings to the gauge
bosons $WW$ and $ZZ$ are unaltered, hence these branching ratios get
suppressed compared to the SM as long as $h b \bar{b}$ is dominant.
For heavy Higgs mass range, $M_h \geq 155$ GeV, the $WW$ mode starts to
dominate, and hence the branching ratio to this mode is very similar
to the SM. The same is true for the $ZZ$ mode. The branching ratio
for the $ZZ$ mode is also essentially the same as the SM for larger mass
ranges ($M_h \geq 185$ GeV).

%\subsection{Higgs productions and signals: implications at the Tevatron}

\begin{center}
\begin{figure}[!t]
\includegraphics[width=3.5in,height=2.3in]{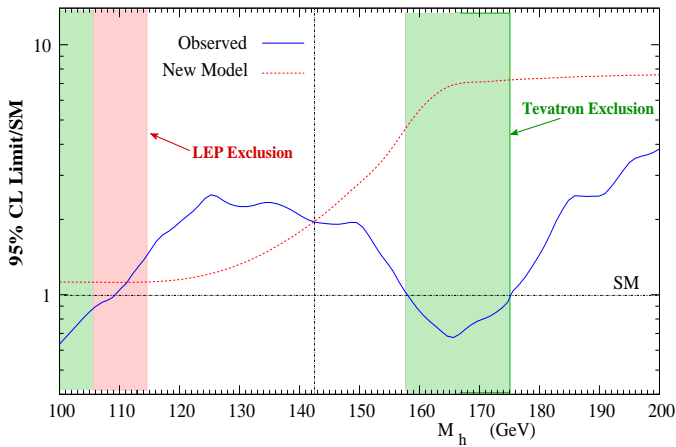}
\caption{Illustrating how the Tevatron bound on SM Higgs applies on the Higgs
boson in our model.}
\label{higgsbound}
\end{figure}
\end{center}

Now we discuss Higgs production and the ensuing final state
signals in our model and contrast those with the SM. First we
consider the Higgs search at the Fermilab Tevatron. For the SM Higgs
boson, recent combined analysis by the CDF and D0 collaborations
(using $6.7\ fb^{-1}$ of data) has excluded the SM Higgs mass range
from $158$ to $175$ GeV at 95\% confidence level (C.L.)
\cite{Aaltonen:2010yv,tev:2010ar}. The dominant production mechanism
for the Higgs boson is gluon gluon fusion via the top quark loop.
Since in our model, the coupling of the Higgs to the top quark is
three times larger, the Higgs production cross sections will be nine
times larger than the SM. Higgs production via the gauge
interactions to $Wh$ and $Zh$ in our model remains the same as in
the SM. Combined Tevatron analysis includes the Higgs signals for
all channels, and the corresponding backgrounds. Their experimental
curve for the observation of the Higgs signals at 95\% C.L. over the
SM expectation  curve as a function of the Higgs mass is shown by
the solid curve in fig. \ref{higgsbound} \cite{tev:2010ar}. The
corresponding SM expectation is shown by the horizontal dash-dotted
line. As shown by the Tevatron analysis (solid curve), the SM Higgs
mass in the range of $158 - 175$ GeV is excluded. The corresponding
exclusion in the low mass range is $M_h \geq 109$ GeV which falls
short of the LEP exclusion of $M_h \geq 114.4$ GeV
\cite{Barate:2003sz}. To apply this combined CDF-D0 analysis to our
model, we have calculated the $\sigma_{p\bar{p}\to h} \times BR
(h\to all$) included by the Tevatron, and compared those with the
SM. The dashed curve in fig. \ref{higgsbound} shows our results for
the ratio of the $\sigma_{p\bar{p}\to h} \times BR (h\to all)$ in
our model to the $\sigma_{p\bar{p}\to h} \times BR (h\to all)$ in
the SM as a function of the Higgs mass. The intersection of the
dashed curve with the solid curve indicates an estimate of the Higgs
mass range ($M_h \gtrsim 142$ GeV) that would be excluded by the
present Tevatron analysis in our model.
%%%%
%\begin{widetext}
%\end{widetext}
%%%%
In the low Higgs mass range, the lower exclusion range increases
slightly from $M_h >109$ GeV in the SM to $M_h >112$ GeV in our
model. As the Tevatron luminosity accumulates further, its increased
sensitivity to our model will help it study a bigger mass range of
the Higgs boson than in the SM. Also, we note that for light Higgs
($M_h < 130$ GeV), the width of the Higgs boson in our model is
larger by a factor of 9 compared to the SM. This can be tested in a
possible future muon or $e^+ e^-$ collider.
%%%%%
%\subsection{Higgs productions and signals: implications for the LHC}

\begin{figure}[!b]
\begin{center}
\includegraphics[width=3.5in,height=2.4in]{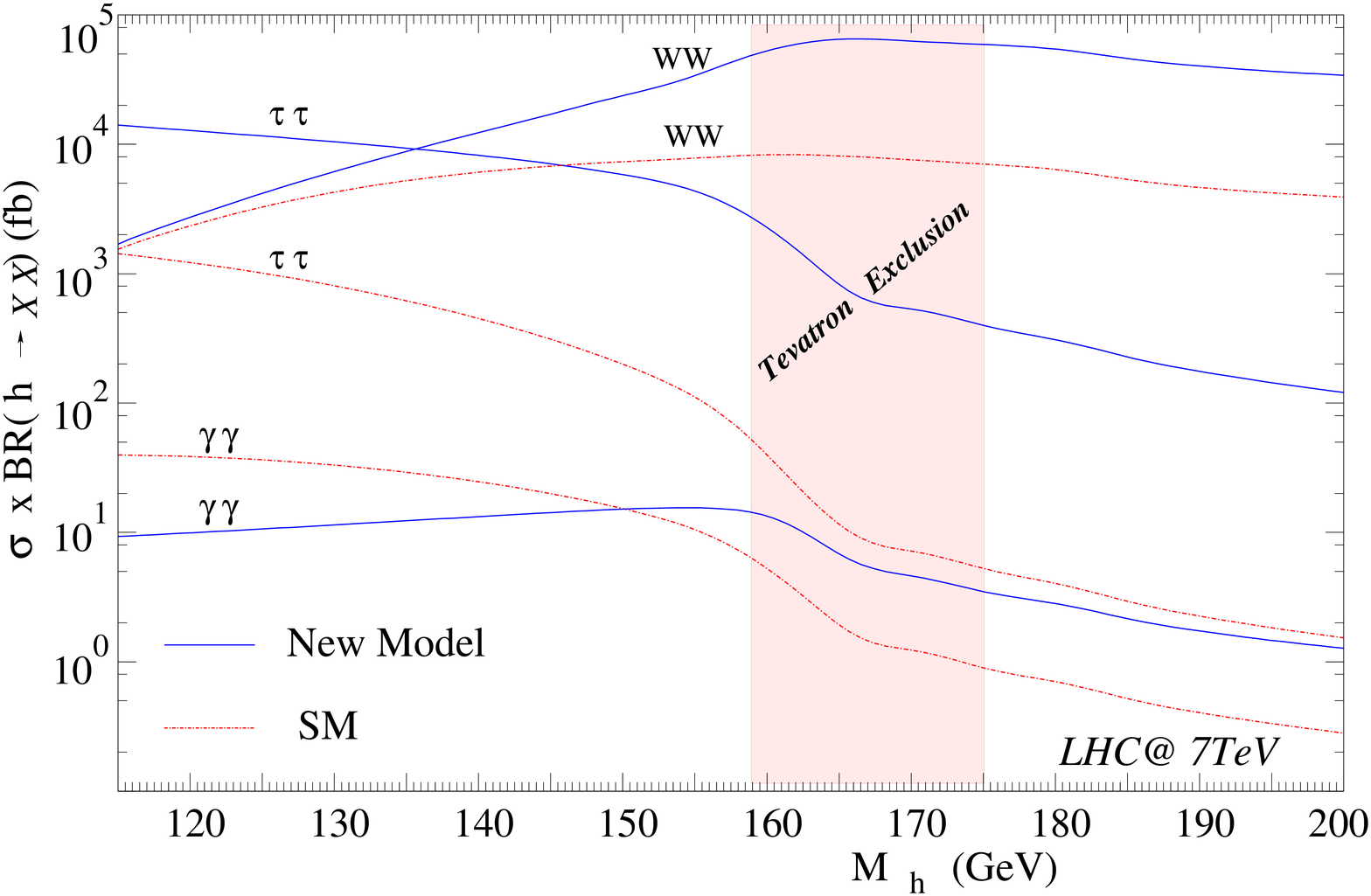}
\end{center}
\caption{Illustrating $\sigma\times BR$ for the SM Higgs and in our model
for the decay modes $\tau\tau,\gamma\gamma$ and $WW$ at LHC with a center-of-mass
energy of 7 TeV.}
\label{sigmaXbr}
\end{figure}

At the LHC, in the SM for large Higgs mass, $M_h > 150$ GeV, the most
promising signals to observe the Higgs boson is via its dominant
production through gluon gluon fusion (or $WW$ fusion), and then
its subsequent decays to $WW$ or $ZZ$. In our model, since the
dominant Higgs productions via gluon gluon fusion is nine times
larger, the Higgs signals will be much stronger. The expectation for
the Higgs signals in few of the relevant modes in our model is
shown in fig. \ref{sigmaXbr} (solid curve), and are compared with the SM
expectations (dash-dotted curves) at the LHC for $\sqrt{s} = 7$ TeV.
Note that the cross section times the branching ratio of $h \rightarrow WW$
in our model is larger than the SM by a factor of $\sim 3 - 9$ for the Higgs
mass range of $150 - 200$ GeV. The same is true for the $ZZ$ mode.
%%%%%%%
%%%%%%
For the low mass range of the Higgs boson, $M_h \sim 115 - 130$ GeV,
the $\gamma \gamma$ mode is the most promising in the SM. In our
model though, as shown in fig. $3$, the signal for the $\gamma \gamma$
mode is reduced by a factor of $\sim 3 - 5$ compared to the
SM. However, the signal in the $\tau \tau$ mode is enhanced almost
by a factor of nine. Thus in our model, signal in the $\tau\tau$
mode may be observable at the LHC for the low Higgs mass range with
good $\tau$ ID for the ATLAS and CMS detectors.

%\section{Other implications}
Inclusion of dimension six operators in the Yukawa sector
%and the Higgs potential (ignoring the dimension 4 operators)
also leads to enhancement in the other modes of Higgs production at
colliders. The associated production of a Higgs boson with a
heavy quark pair ({\sl e.g. $t\bar{t}h$}) is enhanced by a factor of
$9$. The increased event rate would help in improving the
sensitivity for the top-Yukawa coupling in this channel at LHC
\cite{Beenakker:2001rj,Maltoni:2002jr}.

\begin{figure}[!t]
\begin{center}
\includegraphics[width=3.5in,height=2.3in]{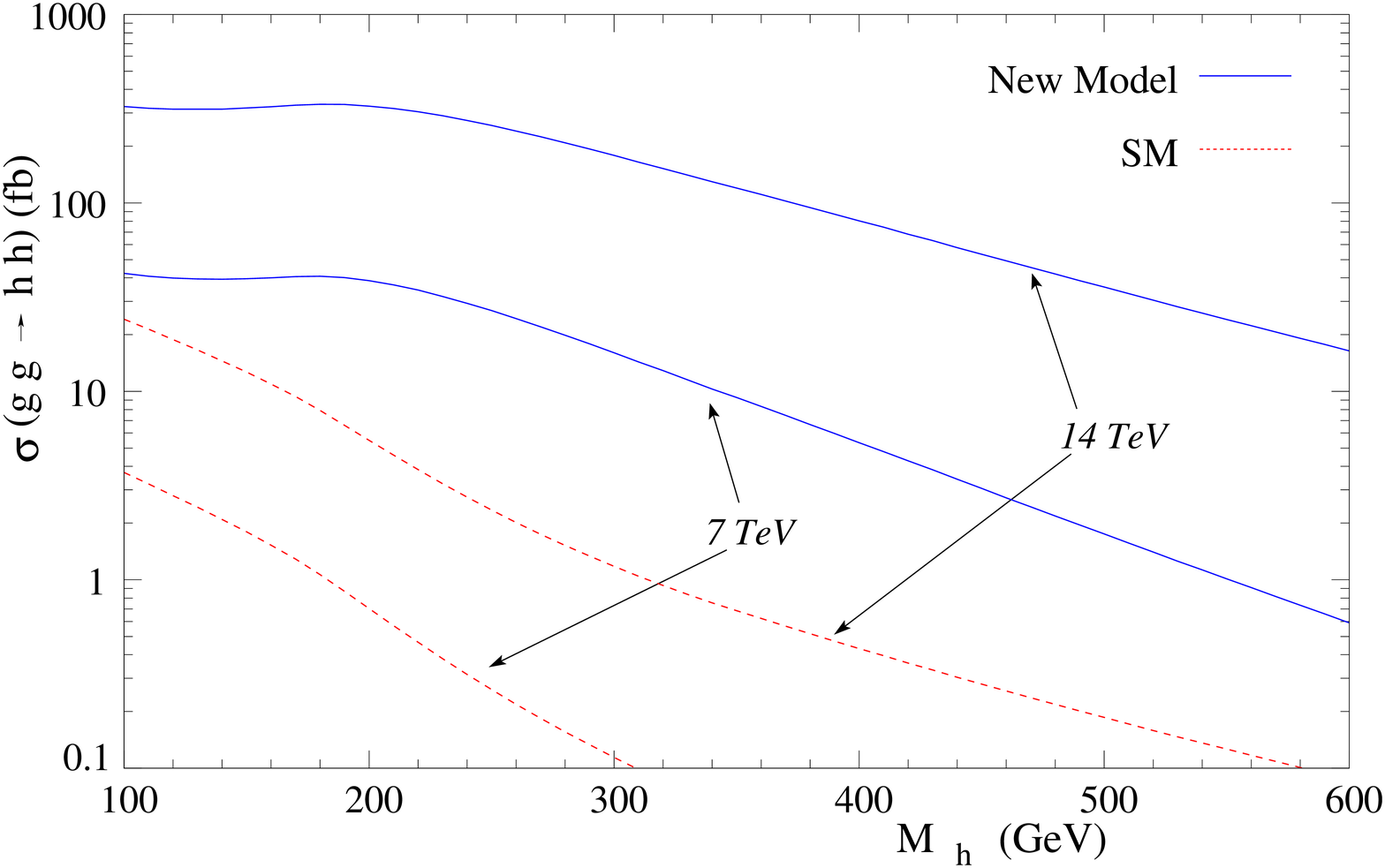}
\end{center}
\caption{Cross section for double Higgs production through gluon gluon fusion
for the SM Higgs (dashed) and for the Higgs in our model (solid)
at LHC with a center-of-mass energy of 7 and 14 TeV.
}
\label{hpair}
\end{figure}

Another important implication of our model is on double Higgs production at the
LHC which can probe the triple Higgs vertex in SM. In the SM,
double Higgs production at LHC proceeds through gluon gluon
fusion at one-loop level through the top quark dominated {\sl
triangle} and {\sl box} diagrams
\cite{Glover:1987nx,Dicus:1987ez,Plehn:1996wb}. Due to additional
contributions coming from the terms involving the dimension six
operators, there is an enhancement in all the vertices involving the
Higgs boson in our model. The box contribution is enhanced by a
factor of 9 in its amplitude because of two Yukawa vertices, while
the triangle contribution is enhanced by a factor of 5, after
combining the new Yukawa and triple Higgs vertices (arising from the Higgs potential where we neglect the dimension 4 operator). There is an
additional contribution to the amplitude through a new interaction
term ($\bar{f}_L f_R h^2$) with a coupling strength of
($\frac{6im_f\alpha_{EW}}{M_W^2}$) where $m_f$ is the mass of the
fermion which leads to a large enhancement of the double Higgs
production cross section at LHC. The analytical formula for the
double Higgs production in SM can be found in
Ref.\cite{Glover:1987nx,Plehn:1996wb}. To put our results in context
we can rewrite the contributions in our model as
%%%%
\begin{align}\label{2hamp}
A_{\triangle}^{NP} &= 5\times A_{\triangle}^{SM} + 2 \times A_{\triangle}^{SM} \frac{\hat{s}-M_h^2}{M_h^2} \nonumber \\
A_{\square}^{NP} &= 9\times A_{\square}^{SM}
\end{align}
%%%
We plot the double Higgs production cross section\footnote{We use the public code
available on M. Spira's webpage (http://people.web.psi.ch/spira/proglist.html)}
as a function of the Higgs mass in fig. \ref{hpair} for both the SM as well as our model.
Although Eq. \ref{2hamp} shows a large enhancement in the
individual contributions, there still is large cancelation between
the box and triangle contributions and so the enhancement in the
cross section compared to the SM is only at the level of a factor of
$\sim 10$ for low Higgs masses as shown in fig. \ref{hpair} which
increases as we go higher in the Higgs mass. Nevertheless it is a
substantial increase for the light Higgs mass range and gives a
cross section of around $\sim 300$ fb at LHC with $\sqrt{s}=14$ TeV
and $\sim 40$ fb with $\sqrt{s}=7$ TeV, respectively for $M_h \leq
220$ GeV. This can give large enough event rates to study the double
Higgs production at LHC.

Finally, let us comment on the scale of new physics, $M$. Up to
dimension six, we can write the Higgs potential as
\begin{align}
\label{pot}
 V_{\text{New}} &= -\mu^2 (H^\dag H) + \lambda (H^\dag H)^2 +
 \frac{1}{M^2} (H^\dag H)^3.
%\nonumber \\
\end{align}
Choosing $\lambda$ to be zero, the condition for the global minima
gives
\begin{align}
\label{seesaw}
 M_h M &= \sqrt{3} v^2.
\end{align}
Using the LEP bound for the Higgs mass, $M_h > 114$ GeV, from
Eq.~\ref{pot},  we obtain $M \leq 1$ TeV. Note the interesting
see-saw type relation between the $M_h$ and $M$ in Eq.~\ref{seesaw}.
Thus if our point of view is correct, we expect the new physics to
appear below the TeV scale.
%%%%%%%
%%%%%%
%%\section*{Acknowledgments}

We are grateful to A. Khanov of the D0 collaboration for many
helpful discussions, especially regarding the combined CDF-D0 Higgs
mass exclusion ranges in the SM and in our new model. This work is
supported in part by the United States Department of Energy, Grant
Numbers DE-FG02-04ER41306 and DE-FG02-04ER46140.

%\vspace{0.2in}
\bibliography{Non-renormalizable Yukawa Interactions and Higgs Physics}% Produces the bibliography via BibTeX.

\begin{thebibliography}{999}


\bibitem{Miransky:1989ds}

%\bibitem{Bardeen:1989ds}
  W.~A.~Bardeen, C.~T.~Hill and M.~Lindner,
  %``Minimal Dynamical Symmetry Breaking Of The Standard Model,''
  Phys.\ Rev.\  D {\bf 41}, 1647 (1990).
  %%CITATION = PHRVA,D41,1647;%%
  V.~A.~Miransky, M.~Tanabashi and K.~Yamawaki,
  %``Dynamical Electroweak Symmetry Breaking with Large Anomalous Dimension and
  %t Quark Condensate,''
  Phys.\ Lett.\  B {\bf 221}, 177 (1989);
  %%CITATION = PHLTA,B221,177;%%
  V.~A.~Miransky, M.~Tanabashi and K.~Yamawaki,
  %``Is the t Quark Responsible for the Mass of W and Z Bosons?,''
  Mod.\ Phys.\ Lett.\  A {\bf 4}, 1043 (1989);
  %%CITATION = MPLAE,A4,1043;%%

\bibitem{Hill:1990ge}
  C.~T.~Hill, M.~A.~Luty and E.~A.~Paschos,
  %``Electroweak symmetry breaking by fourth generation condensates and the
  %neutrino spectrum,''
  Phys.\ Rev.\  D {\bf 43}, 3011 (1991).
  %%CITATION = PHRVA,D43,3011;%%

%\cite{Hill:2002ap}
\bibitem{Hill:2002ap}
  C.~T.~Hill and E.~H.~Simmons,
  %``Strong dynamics and electroweak symmetry breaking,''
  Phys.\ Rept.\  {\bf 381}, 235 (2003)
  [Erratum-ibid.\  {\bf 390}, 553 (2004)];
%  [arXiv:hep-ph/0203079];
  %%CITATION = PRPLC,381,235;%%
%\cite{Contino:2010rs}
%\bibitem{Contino:2010rs}
  R.~Contino,
  %``Tasi 2009 lectures: The Higgs as a Composite Nambu-Goldstone Boson,''
  arXiv:1005.4269 [hep-ph], {\sl and references therein}.
  %%CITATION = ARXIV:1005.4269;%%

\bibitem{Grojean:2004xa}
  C.~Grojean, G.~Servant and J.~D.~Wells,
  %``First-order electroweak phase transition in the standard model with a  low
  %cutoff,''
  Phys.\ Rev.\  D {\bf 71}, 036001 (2005).
%  [arXiv:hep-ph/0407019].
  %%CITATION = PHRVA,D71,036001;%%

%
\bibitem{Barger:2003rs}
  V.~Barger, T.~Han, P.~Langacker, B.~McElrath and P.~Zerwas,
  %``Effects of genuine dimension-six Higgs operators,''
  Phys.\ Rev.\  D {\bf 67}, 115001 (2003).
%  [arXiv:hep-ph/0301097].
  %%CITATION = PHRVA,D67,115001;%%


%\cite{Buchmuller:1985jz}
\bibitem{Buchmuller:1985jz}
  W.~Buchmuller and D.~Wyler,
  %``Effective Lagrangian Analysis Of New Interactions And Flavor
  %Conservation,''
  Nucl.\ Phys.\  B {\bf 268}, 621 (1986);
  %%CITATION = NUPHA,B268,621;%%
%\cite{Grzadkowski:2010es}
%\bibitem{Grzadkowski:2010es}
  B.~Grzadkowski, M.~Iskrzynski, M.~Misiak and J.~Rosiek,
  %``Dimension-Six Terms in the Standard Model Lagrangian,''
  arXiv:1008.4884 [hep-ph].
  %%CITATION = ARXIV:1008.4884;%%

\bibitem{hdecay}
  M.~Spira,
  %``HIGLU and HDECAY: Programs for Higgs boson production at the LHC and  Higgs
  %boson decay widths,''
  Nucl.\ Instrum.\ Meth.\  A {\bf 389}, 357 (1997);
 % [arXiv:hep-ph/9610350];
  %%CITATION = NUIMA,A389,357;%%
%\cite{Djouadi:1997yw}
%\bibitem{Djouadi:1997yw}
  A.~Djouadi, J.~Kalinowski and M.~Spira,
  %``HDECAY: A program for Higgs boson decays in the standard model and its
  %supersymmetric extension,''
  Comput.\ Phys.\ Commun.\  {\bf 108}, 56 (1998).
 % [arXiv:hep-ph/9704448].
  %%CITATION = CPHCB,108,56;%%

%\cite{Aaltonen:2010yv}
\bibitem{Aaltonen:2010yv}
  T.~Aaltonen {\it et al.}  [CDF and D0 Collaborations],
  %``Combination of Tevatron searches for the standard model Higgs boson in the
  %W+W- decay mode,''
  Phys.\ Rev.\ Lett.\  {\bf 104}, 061802 (2010).
  %[arXiv:1001.4162 [hep-ex]].
  %%CITATION = PRLTA,104,061802;%%

%\cite{tev:2010ar}
\bibitem{tev:2010ar}
    [The TEVNPH Working Group of the CDF and D0 Collaborations],
  %``Combined CDF and D0 Upper Limits on Standard Model Higgs-Boson Production
  %with up to 6.7 fb$^{-1}$ of Data,''
  arXiv:1007.4587 [hep-ex].
  %%CITATION = ARXIV:1007.4587;%%

%\cite{Barate:2003sz}
\bibitem{Barate:2003sz}
  R.~Barate {\it et al.}
%[LEP Working Group for Higgs boson searches an ALEPH Collaboration],
  %``Search for the standard model Higgs boson at LEP,''
  Phys.\ Lett.\  B {\bf 565}, 61 (2003).
%  [arXiv:hep-ex/0306033].
  %%CITATION = PHLTA,B565,61;%%

%\cite{Beenakker:2001rj}
\bibitem{Beenakker:2001rj}
  W.~Beenakker, S.~Dittmaier, M.~Kramer, B.~Plumper, M.~Spira and P.~M.~Zerwas,
  %``Higgs radiation off top quarks at the Tevatron and the LHC,''
  Phys.\ Rev.\ Lett.\  {\bf 87}, 201805 (2001).
%  [arXiv:hep-ph/0107081].
  %%CITATION = PRLTA,87,201805;%%

%\cite{Maltoni:2002jr}
\bibitem{Maltoni:2002jr}
  F.~Maltoni, D.~L.~Rainwater and S.~Willenbrock,
  %``Measuring the top-quark Yukawa coupling at hadron colliders via  t anti-t
  %h, h --> W+ W-,''
  Phys.\ Rev.\  D {\bf 66}, 034022 (2002).
 % [arXiv:hep-ph/0202205].
  %%CITATION = PHRVA,D66,034022;%%

 \bibitem{Dicus:1987ez}
  D.~A.~Dicus, C.~Kao and S.~S.~D.~Willenbrock,
  %``HIGGS BOSON PAIR PRODUCTION IN THE EFFECTIVE W APPROXIMATION,''
  Phys.\ Rev.\  D {\bf 38}:1088 (1988).
  %%CITATION = PHLTA,B200,187;%%

%\cite{Glover:1987nx}
\bibitem{Glover:1987nx}
  E.~W.~N.~Glover and J.~J.~van der Bij,
  %``HIGGS BOSON PAIR PRODUCTION VIA GLUON FUSION,''
  Nucl.\ Phys.\  B {\bf 309}, 282 (1988).
  %%CITATION = NUPHA,B309,282;%%

%\cite{Plehn:1996wb}
\bibitem{Plehn:1996wb}
  T.~Plehn, M.~Spira and P.~M.~Zerwas,
  %``Pair Production of Neutral Higgs Particles in Gluon--Gluon Collisions,''
  Nucl.\ Phys.\  B {\bf 479}, 46 (1996)
  [Erratum-ibid.\  B {\bf 531}, 655 (1998)].
 % [arXiv:hep-ph/9603205].
  %%CITATION = NUPHA,B479,46;%%

%\cite{Asakawa:2010xj}
%\bibitem{Asakawa:2010xj}
%  E.~Asakawa, D.~Harada, S.~Kanemura, Y.~Okada and K.~Tsumura,
  %``Higgs boson pair production in new physics models at hadron, lepton, and
  %photon colliders,''
%  arXiv:1009.4670 [hep-ph].
  %%CITATION = ARXIV:1009.4670;%%


\end{thebibliography}

\end{document}